\documentclass[aps,prl,dvips,twocolumn]{revtex4}

\usepackage{graphicx}
\def\be{\begin{equation}}
\def\ee{\end{equation}}

\begin{document}

\title{Ab initio theory of coherent phonon generation by laser excitation}

\author{Y. Shinohara}
\affiliation{Graduate School of Science and Technology, University of Tsukuba,
Tsukuba 305-8571, Japan}
\author{K. Yabana}
\affiliation{Graduate School of Science and Technology, University of Tsukuba,
Tsukuba 305-8571, Japan}
\affiliation{Center for Computational Sciences, University of Tsukuba,
Tsukuba 305-8571, Japan}
\author{Y. Kawashita}
\affiliation{Graduate School of Science and Technology, University of Tsukuba,
Tsukuba 305-8571, Japan}
\author{J.-I. Iwata}
\affiliation{Center for Computational Sciences, University of Tsukuba,
Tsukuba 305-8571, Japan}
\author{T. Obote}
\affiliation{
Advanced Photon Research Center, Japan Atomic Energy Agency, Kizugawa, Kyoto
619-0215, Japan}
\author{G.F. Bertsch}
\affiliation{Institute for Nuclear Theory and Dept. of Physics,
University of Washington, Seattle, Washington}

\begin{abstract}

We show that time-dependent density functional theory (TDDFT) is
applicable to coherent optical phonon generation by intense laser
pulses in solids. The two mechanisms invoked in phenomenological theories, 
namely
impulsively stimulated Raman scattering and displacive excitation, are
present in the TDDFT. Taking the example of crystalline Si, we find
that the theory reproduces the phenomena observed 
experimentally: dependence on polarization, strong growth at the direct 
band gap, and the change of phase from below to above the band gap. 
We conclude that the
TDDFT offers a predictive {\it ab initio} framework 
to treat coherent optical phonon generation.  
\end{abstract}

\maketitle

%\subsection{Introduction}

There has been much experimental progress in the study of 
intense electromagnetic fields interacting with condensed
matter using pump-probe techniques on femtosecond time
scales \cite{ro02}. These interactions are a challenging 
subject for theory, in view of the need to go beyond 
perturbative methods in dealing with strong fields.
One promising theoretical approach useful to describe electron 
dynamics on femtosecond time scales is time-dependent density 
functional theory (TDDFT) \cite{rg84}.  In this Letter we apply the TDDFT 
to the generation of coherent phonons by strong laser pulses. Our
goals are both to test the utility of the TDDFT in this domain
and to assess the validity of phenomenological models that are in current use.
In the past, two mechanisms have been invoked to explain the 
generation of coherent phonon \cite{me97,st02}. The impulsively stimulated 
Raman scattering was proposed for the coherent phonon 
generation in dielectrics with a laser pulse whose frequency 
is lower than the direct band gap. 
In this mechanism, electrons are virtually excited
following adiabatically the laser electric field.
The crucial quantity is the Raman tensor, the derivative
of dielectric function with respect to the phonon coordinate.
The other mechanism, called displacive
excitation,  requires higher frequencies to generate real electron-hole 
excitations in the final state \cite{ze92,sc93,ku94}.  
These excitations then shift the equilibrium
position of the phonon coordinates. 
In this work we consider a bulk Si under irradiation of laser pulses of 
frequencies below and above the direct band gap, and show
that the TDDFT is computationally feasible, includes two
above-mentioned mechanisms, and produces results that are 
in qualitative agreement with experiments \cite{ha03,ri07}. 
The TDDFT calculations prove to be also useful to evaluate 
phenomenological and macroscopic models for the phonon 
generation process.

Our computational framework is based on equations of motion
derived from a Lagrangian for a periodic crystalline system
under a time-dependent, spatially uniform electric field \cite{biry}.
The Lagrangian is 
\begin{eqnarray}
\label{L}
L &=& \sum_i\int_{\Omega} d\vec r
 \left\{ 
 \psi_i^* i\frac{\partial }{\partial t}\psi_i
-\frac{1}{2m}
\left\vert\left( -i\vec \nabla+
\frac{e}{c} \vec A \right) \psi_i \right\vert^2\right\}
\nonumber\\
&& -\int_\Omega d \vec r \left\{(e n_{ion} - e n_e) \phi - E_{xc}[n_e]\right\}
\nonumber\\
&& +\frac{1}{8\pi} \int_\Omega d \vec r (\vec \nabla \phi)^2
+\frac{\Omega}{8\pi c^2}
\left( \frac{d \vec A}{dt} \right)^2
\nonumber\\
&& + \frac{1}{2}\sum_{\alpha} M_{\alpha} \left( \frac{d\vec R_{\alpha}}{dt} \right)^2
+\frac{1}{c}\sum_\alpha Z_\alpha e \frac{d \vec R_\alpha}{dt}\vec A\,.
\end{eqnarray}
Here $\psi_i$ is the time-dependent electron orbitals, 
taken as Bloch orbitals in a unit cell of volume $\Omega$.
$n_e(\vec r,t) = \sum_i \vert \psi_i(\vec r,t) \vert^2$ 
represents the electron density distribution.
$\vec R_{\alpha}$ are atomic positions.
The electromagnetic field terms are split into a long-range
spatially uniform part $\vec A(t)$ and a periodic part given
by a Coulomb potential $\phi$.
Variations with respect to the orbitals $\psi_i$,
potential $\phi$, and atomic coordinates $\vec R_{\alpha}$
result in the time-dependent Kohn-Sham equation for
$\psi_i$, the Poisson equation for $\phi$, and the
Newton equation for $\vec R_{\alpha}$, respectively.
All the equations except those for $\vec R_{\alpha}$
are the same as those employed in \cite{biry} and \cite{ot08}.

To introduce the external laser field, we express the vector
potential $\vec A(t)$ as a sum of an external field 
$\vec A_{\rm ext}(t)$ and the induced field $\vec A_{\rm ind}(t)$, 
with $
\vec A(t) = \vec A_{\rm ext}(t) + \vec A_{\rm ind}(t)
$
and treat $\vec A_{\rm ind}(t)$ as dynamic.
The variation with respect to $\vec A_{\rm ind}(t)$ yields
the following equation of motion,
\be
\label{A-induced}
\frac{\Omega}{4\pi c^2} \frac{d^2 \vec A_{ind}(t)}{dt^2}
= \frac{e}{c}\int_{\Omega} d\vec r
\left\{ \vec j_{ion} - \vec j_e \right\}
-\frac{e^2}{mc^2} N_e \vec A(t)
\ee
To simulate the time-dependent electric field of the
laser pulse, we take $\vec A_{\rm ext}(t)$ to have the form
\be
\label{Aex}
\vec A_{\rm ext} = \int^t dt' {\cal E}_0 \sin^2 \left({\pi t'\over T_p}\right) 
\,\sin \omega t' 
\ee
for $0<t<T_p$ and zero otherwise, with $T_p=16$ fs and ${\cal E}_0$
corresponding to peak intensity 
$I=10^{12}$ W/cm$^2$.

\begin{figure}
\includegraphics [width = 5cm]{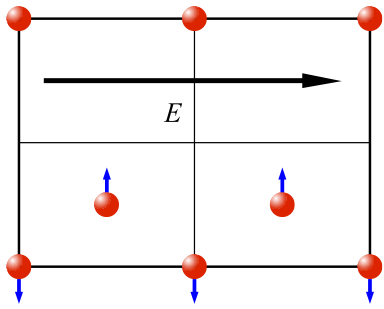}
\caption{\label{geometry}
Geometry of the electric field and the optical phonon 
displacement in the 8-atom unit cell. 
The $[011] \times [100]$ plane and atoms on the plane are
drawn with small arrows which show the direction of the
optical phonon coordinate.
}
\end{figure}

The laser pulse is directed on a $[100]$ Si 
surface at normal incidence with a linear polarization 
oriented along the $[011]$ axis.
We show in Fig. \ref{geometry} the atomic positions of 
Si atoms in the plane defined by the $[011]$ and $[100]$ axes.
The 4 atoms lying on the plane are shown. The optical phonon
coordinate which couples to the laser field is  
shown by vertical blue arrows.

\begin{figure}
\includegraphics [width=6cm]{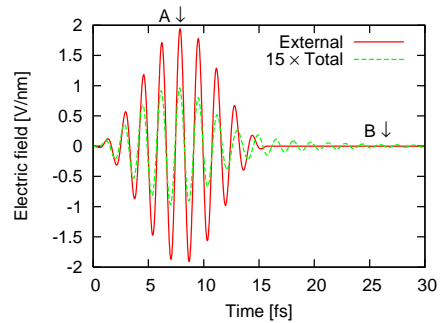}
\caption{\label{Eext_tot} 
Electric fields are shown as a function of time.
The red solid line shows the applied laser pulse, 
Eq. (\ref{Aex}), characterized by the peak intensity, 
$I=10^{12}$W/cm$^2$, frequency $\hbar \omega=2.5$ eV, 
and the pulse duration, $T_p=16$~fs.
The green dashed line shows the summed electric field of
applied and induced ones, multiplied by a factor 15.
}
\end{figure}

\begin{figure}
\includegraphics [width=5cm]{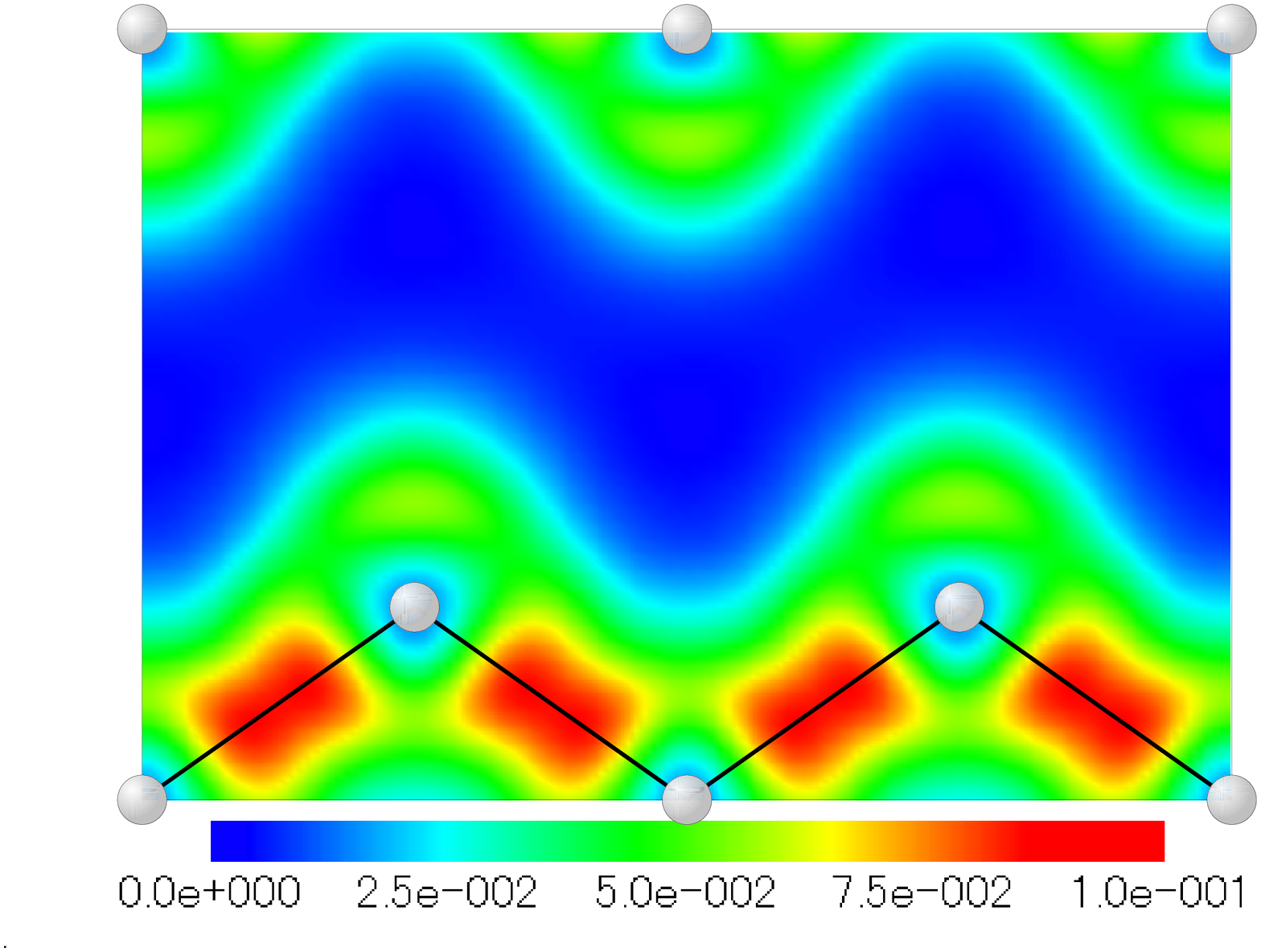}
\includegraphics [width=5cm]{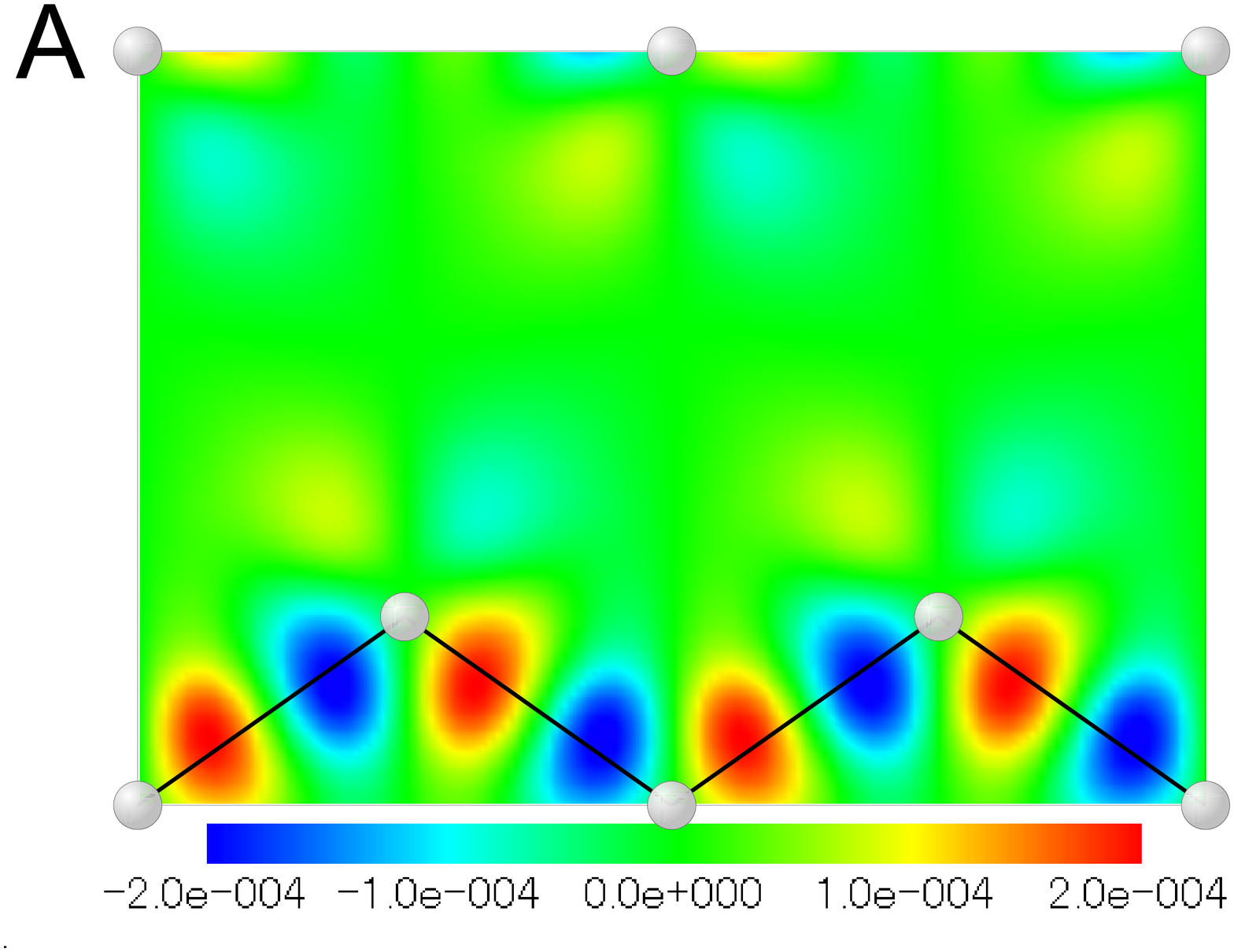}
\includegraphics [width=5cm]{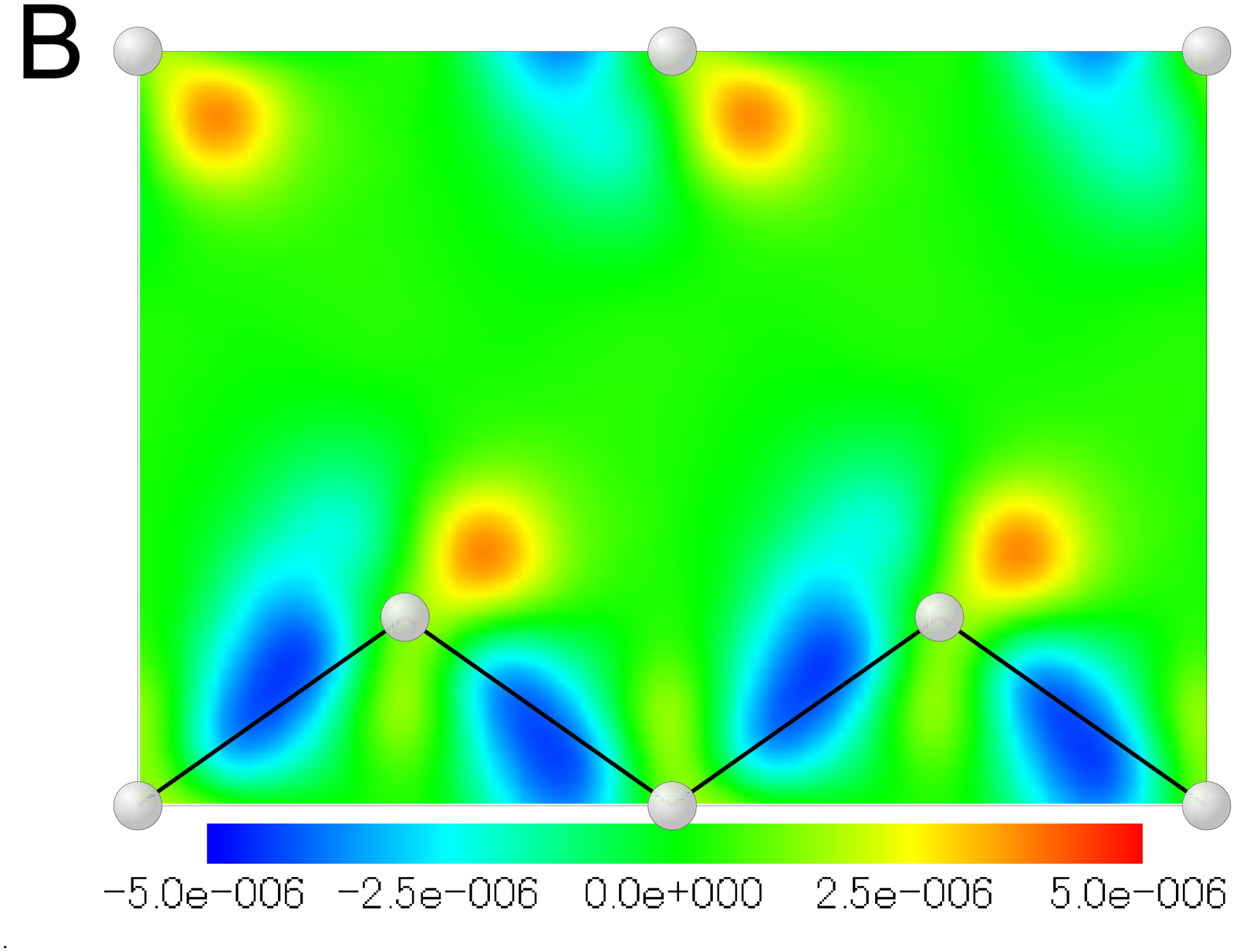}
\caption{\label{e-dns} 
Top panel shows the ground-state electron density in the 
plane shown in Fig. \ref{geometry}.
The middle and bottom panels show the change of the 
electron density from that in the ground state by
the laser pulse described in Fig. \ref{Eext_tot}.
The middle panel corresponds to the time $A$ and the
bottom panel to the time $B$ in Fig. \ref{Eext_tot},
respectively. In the middle and bottom panels,
the red color indicates the increase of the electron 
density, while blue color indicates the decrease.
}
\end{figure}

%\subsection{Numerical details}

Our calculations are based on the LDA density functional \cite{pz81},
treating the four valence electrons of Silicon explicitly and using
the Troullier-Martins pseudopotential \cite{TM}. 
We employ the real-time and real-space scheme which was developed
by us \cite{yb96}. The geometry is taken to be a simple cubic unit cell
containing 8 Si atoms, with lattice constant $a=10.26$ au.
We have carefully examined the convergence of the results with
respect to numerical parameters. We find that a spatial division of 
$16^3$, $k$-space grid of $24^3$, and the time step of $\Delta t=0.08$ au
is adequate for our purposes, and these numerical parameters are 
adopted for the results reported below. 
To make the present calculation feasible, parallel computation
distributing $k$-points into processors is indispensable.
We note the calculated direct band gap of Si is 2.4 eV, smaller than
the measured value of 3.3 eV.

%\subsection{Electron Dynamics}

We first show the electron dynamics induced by a laser pulse. 
Figure \ref{Eext_tot} shows the time dependence of the 
electric fields. The red solid curve shows the electric field of 
applied laser pulse $E_{\rm ext}(t)=-(1/c)dA_{\rm ext}/dt$. 
We choose the laser frequency $\hbar \omega = 2.5$ eV, close 
to the value of the direct band gap. The green dashed curve shows 
the sum of the applied and induced electric fields, 
$E_{\rm tot}(t) = E_{\rm ext}(t)+ E_{\rm ind}(t)$. 
The difference of the magnitudes of the two fields comes 
from a dielectric screening.

Figure \ref{e-dns} shows the electron density in the plane
of Fig. \ref{geometry}. The top panel shows the ground-state 
electron density, and the middle and bottom panels show the 
change of electron density from that in the ground state at 
two times, marked  $A$ and $B$ in Fig. \ref{Eext_tot}, respectively.
In the middle and bottom panels, red and blue indicate an increase 
or decrease of electron density, respectively.
At time $A$, the electric field is maximum
and there is a strong virtual excitation of the electrons.
In the middle panel of Fig. \ref{e-dns}, a movement of electrons
is seen in the bond connecting two Si atoms.
At the time $B$, the external electric field ended.
Since the ultrashort laser pulse includes frequency components
above the direct band gap, there appear real electron-hole
excitations. In the bottom panel of Fig. \ref{e-dns}, one can see that
the excitation results in a decreased density in the bond
region and an increase near the Si atoms but away from the bond.
One should note that the coloring
of the middle and bottom figures are different by a factor
of 40 to improve the visibility of the density change at time $B$.

\begin{figure}
\includegraphics [scale=1.0] {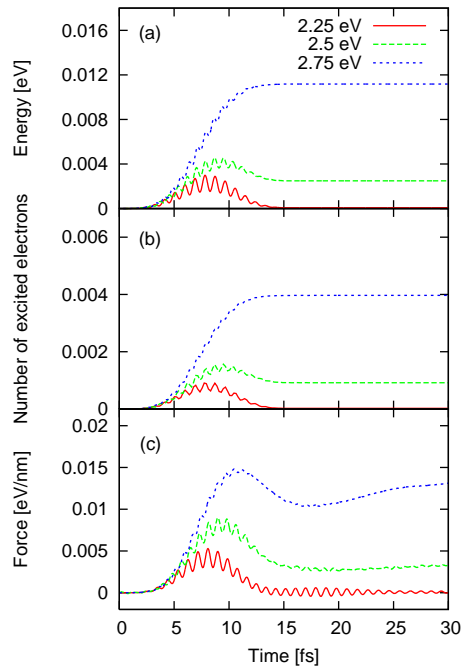}
\caption{\label{Eex} 
Electron excitation of the crystal during and after
the pulse for several laser frequencies across
the direct band gap.  The top panel (a) shows the 
energy in the unit cell including electron-hole 
excitation energy and the electric field energy.  
The middle panel (b) shows the the number of 
electron-hole pairs in the unit cell.
The bottom panel (c) shows the force on the
optical phonon coordinate.
}

\end{figure}

We next examine how the character of the electronic 
excitation changes as the laser frequency increases 
from below to above the direct band gap.
Characteristics of the excitation as a function of time
are shown for frequencies $\hbar \omega = 2.25$ eV, 
2.5 eV, and 2.75 eV in Fig. \ref{Eex}.  The top panel
shows the total increase in energy in the unit cell, including
both electronic excitation energy and the electromagnetic
field energy.  The red solid curve shows the results for a frequency
below the band gap.  Here the energy drops almost to zero after
the pulse is over, as to be expected.  The green dashed curve,
corresponding to a frequency at the band gap, shows that
some excitation energy remains after the end of the pulse,
comparable in magnitude to the total energy at the peak.
Finally, the blue dotted curve shows that above the gap the
laser-electron interaction is highly dissipative, leaving
a large excitation energy in the final state.
The middle panel in the figure shows the number of 
excited electrons as a function of time.  This is
calculated by taking the overlaps of the time-dependent
occupied orbitals with the initial state static orbitals
as in Ref. \cite{ot08}.  The results are qualitatively 
very similar to what we found for the energy.
Below the direct band gap,
the excited electron shows a peak during the pulse and 
then drops off to a very small value in the final state.  
At higher frequencies, 
the excitations remain in the final state and it
is not possible to distinguish the real excitation from the 
virtual one during the pulse.  In summary, one sees an
adiabatic response below the gap switching rather
abruptly to a strongly dissipative response above
the gap.

Finally, in the bottom figure, we show the calculated induced 
force for the three frequencies.  Note that the ion positions 
are fixed in these calculations; the accelerations are small 
and the resulting displacements would be inconsequential.
The lowest frequency, shown by the red solid curve, gives 
a force envelope that follows the shape of the pulse intensity.
This is just what one would expect from the adiabatic formula~
\cite[Eq. (2)]{me97}.  
One also sees high frequency oscillations superimposed on the 
envelope of the curve.  The frequency of these oscillations are 
twice the laser frequency, again as expected from the adiabatic formula.
The green dashed curve shows the force for a laser frequency of
$\hbar \omega = 2.5$, nearly at the direct
band gap.  One still sees a large peak at 10 fs  associated with
instantaneous high field intensity.  However, there is a residual force
after the end of the pulse which is rather constant with time.
This is just what one expects for displacive mechanism.  At this
point, we have shown that TDDFT reproduces at a qualitative level the
role of the two mechanisms.   Beyond that, the relative 
sign associated with them can be extracted from the graph.
The last case shown, $\hbar \omega = 2.75$, is $0.35$ eV above the
direct gap.  Here the displacive mechanism is completely dominant,
although one can still see an enhancement of the force during the
pulse.   

We now integrate the time-dependent force to get the lattice
distortion associated with the phonon coordinate.  In principle,
the restoring potential for the lattice vibration is included
in the evolution equations, but the amplitude of the lattice 
displacement is too small numerically to include it in the direct 
integration. So for this part of the analysis we simply assume 
a harmonic restoring potential consistent with the observed 
optical phonon frequency, $f_{phonon} = 15.3 $ THz.

To analyze the characteristics of the coherent phonon,
we fit the oscillation of the displacement to a cosine function
as in conventional parametrization of the experimental reflectivity
measurements \cite{ha03}, 
\be
\label{osc}
q(t) = -q_0 \cos( \omega_{ph} t+ \phi) + \bar q\,.
\ee

\begin{figure}
\includegraphics [scale=1.0] {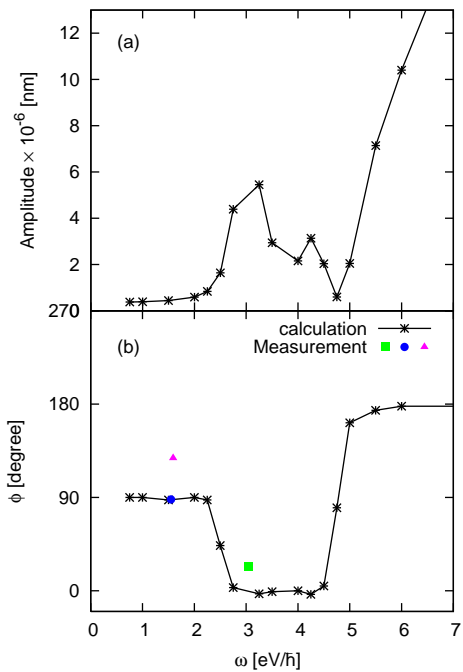}
\caption{\label{amp.4} The amplitude (a) and the phase (b) 
of the phonon oscillation Eq. (\ref{osc}) as a function of 
laser frequency  $\omega$. 
}
\end{figure}

Fig. \ref{amp.4} shows the amplitude and phase as a function
of laser frequency, fitted in the time interval 40-90 fs. 
Below the direct gap energy the phase is close to 
$\pi/2$ as expected for the Raman mechanism.  The amplitude
remains almost constant in this frequency region, also consistent 
with the Raman mechanism.
One sees a quite sharp drop 
from that value to $\phi=0$ as the direct gap is crossed, showing the
transition to the displacive behavior.  The amplitude also shows a
sudden increase across the direct gap.
Several experimental measurements are also shown on the figure for 
the phase.  Two of them  \cite{ha03,ka09} are in the Raman regime.  
The theory supports the results of Ref. \cite{ha03}, which reports 
a value close to $\pi/2$.  The other measurement does not appear 
consistent with our theory or indeed with the other experiment.  
The phase has also been measured in the gap region \cite{ri07}, shown by
the square on Fig. \ref{amp.4}(b).  This point should be compared 
with the theory at the corresponding calculated gap energy, 2.4 eV.  
In both theory and experiment the phase has decreased from the 
Raman value, but decrease seems larger for the experimental measurement. 
Both results are in a range where the mechanism is changing rapidly.  
All in all, we find the agreement quite satisfactory on a qualitative 
level, particularly since the phase could have come out with an 
opposite sign ($\phi\approx \pi$). 

At higher frequencies, the theoretical phase goes to zero as expected 
for the displacive mechanism, but then it rises again beyond 4.5 eV,
approaching $\pi$ at $\hbar \omega = 5$ eV. There is a corresponding
dip and growth in the amplitudes associated with a change in the
sign of the displacive force.
Different electron orbitals are excited at 
the high frequency, and apparently those orbitals have an opposite sign 
contribution to the displacive shift.  

%\subsection{Laser intensity dependence}

We also examined the dependence of amplitude and phase of the coherent
phonon on the intensity of the laser pulse.
At all frequencies we examined, the amplitude
of the phonon is proportional to the laser intensity.
In the impulsive Raman mechanism which is applicable below the band gap,
this dependence is expected from the adiabatic formula \cite[Eq. (2)]{me97}.
In the displacive mechanism, it is also expected if the medium is
excited by a one-photon absorption process.
The phase of the coherent phonon is found to be sensitive only on
the frequency but not on the intensity until the multiphoton
absorption processes become significant.

%\subsection{Summary}

In summary, we have derived and carried out a computational method
to apply time-dependent density functional theory to laser-lattice
interactions, taking as an example the excitation of coherent optical
phonon by femtosecond-scale laser pulse in silicon.
The  qualitative agreement between theory and measured phase
of the coherent phonon confirms the utility of the TDDFT to describe
electron dynamics resulting from intense laser pulse in solids.

%\section*{Acknowledgment}

The numerical calculation were performed on the massively parallel cluster
T2K-Tsukuba, University of Tsukuba, and the supercomputer at the Institute
of Solid State Physics, University of Tokyo.
TO acknowledges support by the Grant-in-Aid for Scientific Research 
No. 21740303. 
GFB acknowledges support by the
National Science Foundation under Grant PHY-0835543 and by the 
DOE under grant DE-FG02-00ER41132.

\end{document}